\newif\iffigs\figstrue
\documentclass[12pt]{article}
\usepackage{latexsym,amssymb}
   \input{epsf}
\else
\fi

\newcommand{\sect}[1]{\setcounter{equation}{0}\section{#1}}

\newcommand{\app}[1]{\setcounter{section}{0}\setcounter{equation}{0}
\renewcommand{\thesection}{\Alph{section}}
\section{#1}}

\newcommand{\eq}{\begin{equation}}
\newcommand{\eqa}{\begin{eqnarray}}
\newcommand{\en}{\end{equation}}
\newcommand{\ena}{\end{eqnarray}}
\newcommand{\enn}{\nonumber \end{equation}}

\def\sk{\vskip .4cm}
\def\noi{\noindent}

\def\al{\alpha}

\def\be{\beta}

\def\Ga{\Gamma}

\def\rhop{\rho'}

\def\we{\wedge}

\def\de{\delta}

\def\part{\partial}

\def\Linv#1#2{ \La^{-1~#1}_{~~~~~#2} }

\def\La{\Lambda}

\def\n2{{{N+1} \over 2}}

\def\square{{\,\lower0.9pt\vbox{\hrule \hbox{\vrule height 0.2 cm
\hskip 0.2 cm \vrule height 0.2 cm}\hrule}\,}}

\def\Q.E.D.{\rightline{$\Box$}}

\def\sumong{\sum_{g \in G}}

\def\sumongnote{\sum_{g \not= e}}

\def\sumonh{\sum_{h \in G}}

\def\Lcal{{\cal L}}
\def\Rcal{{\cal R}}

\def\lb{\label}

\newcommand{\matc}{\begin{array}{c}}
\newcommand{\matcc}{\begin{array}{cc}}
\newcommand{\matccc}{\begin{array}{ccc}}
\newcommand{\matcccc}{\begin{array}{cccc}}
\newcommand{\emat}{\end{array}}

\begin{document}
\begin{titlepage}
\vskip -1cm \rightline{DFTT-37/02} \rightline{SISSA 76/2002/FM}
\rightline{November 2002}
 \vskip 1em
\begin{center}
{\Large\bf Noncommutative de Rham cohomology\\ of finite groups }
\\[3em]
 {\bf L. Castellani}${}^{1,2,3}$ , {\bf
R. Catenacci }${}^{1,3}$, {\bf M. Debernardi} ${}^1$ and {\bf C.
Pagani}${}^4$\\[3em] {\sl ${}^1$ Dipartimento di Scienze,
 Universit\`a del Piemonte Orientale,\\Corso Borsalino 46, Alessandria,
 Italy}
  \\[.5em]{\sl ${}^2$ Dipartimento di Fisica Teorica,
Via P. Giuria 1, 10125 Torino, Italy.}
 \\[.5em]
  {\sl ${}^3$ I.N.F.N., Torino and Alessandria}
    \\[.5em]
   {\sl  ${}^4$  S.I.S.S.A, via Beirut 2-4, 34014 Trieste, Italy}
\end{center}
 \vskip 2 cm
\begin{abstract}
\sk
 We study de Rham cohomology for various differential calculi
on finite groups $G$ up to order 8. These include the permutation
group $S_3$, the dihedral group $D_4$ and the quaternion group
$Q$. Poincar\'e duality holds in every case, and under some
assumptions (essentially the existence of a top form) we find that
it must hold in general.

A short review of the bicovariant (noncommutative) differential
calculus on finite $G$ is given for selfconsistency. Exterior
derivative, exterior product, metric, Hodge dual, connections,
torsion, curvature, and biinvariant integration can be defined
algebraically. A projector decomposition of the braiding operator
is found, and used in constructing the projector on the space of
2-forms. By means of the braiding operator and the metric a knot
invariant is defined for any finite group.

\end{abstract}

\vskip 2 cm

 \leftline{\small ~~~castellani@to.infn.it}
 \leftline{\small~~~catenacc@mfn.unipmn.it}
 \leftline{\small ~~~pagani@sissa.it}

 \vskip .2cm

 \noi \hrule \vskip .2cm

 \noi {\footnotesize
 Supported in part by the European Commission RTN programme
 HPRN-CT-2000-00131 and by MIUR under contract 2001-025492}

\end{titlepage}
\newpage
\setcounter{page}{1}


\sect{Introduction}


Most differential geometric objects pertaining to smooth manifolds
can be generalized in the case of discrete sets. When these sets
are related to a group structure (as for example finite group sets
$G$), the induced Hopf algebra structure on the functionals
$Fun(G)$ gives a canonical way to construct bicovariant calculi on
them \cite{wor}.

Differential geometry plays a basic role in the construction of
field theories describing the fundamental interactions in nature:
gravity and Yang-Mills actions are rooted in Riemannian and fiber
bundle geometry. The idea of translating the concepts of metric,
connection, curvature to discrete cases has been explored in the
past, one of the first fruitful instances being Regge calculus
(for recent reviews and reference lists see for example
\cite{revgravdiscrete}). The physical motivations of this idea
reside in the nonrenormalizability of Einstein gravity (whereas a
field theory on discrete spacetime has no ultraviolet divergences)
and computational advantages in the study of nonperturbative
phenomena in quantum gauge theories by numerical evaluation of
path-integrals. Moreover the possibility of a discrete spacetime,
with a ``granularity" of the order of the Planck length, has
emerged also within the framework of string/brane theories.

Another (and related) approach to the ``algebrization" of geometry
has been pioneered by A. Connes, in the context of noncommutative
geometry \cite{NCGreviews}.

Using the general results of Woronowicz \cite{wor},
(noncommutative) differential calculi have been constructed on
$Fun(quantum~groups)$ (for an introductory review see for ex.
\cite{ACintro,Athesis}) and $Fun(finite~groups)$
\cite{DMGcalculus,FI1,gravfg,tmr,majidetal,CasPag,ACI,DMgrouplattice},
two particular examples of Hopf algebras, of interest for physical
applications. The corresponding differential geometric objects and
operations can be used to construct actions invariant under the
quantum group transformations (see for ex. \cite{LCqgravqgauge}),
or under finite group transformations
\cite{DMgaugegrav,gravfg,tmr,majidetal,CasPag,ACI}, generalizing
the usual gravity and gauge actions.

 On finite
groups $G$ the noncommutativity is mild, in the sense that
functions on $G$ commute between themselves, and only the
commutations between functions and differentials, and of
differentials between themselves are nontrivial.

For smooth manifolds, de Rham cohomology provides a bridge between
differential geometry and topology. It is natural to ask whether
this bridge exists also in the case of finite group manifolds.
Using integration on finite groups, can one define the analogue of
characteristic classes, and relate them to topological properties
of finite group spaces ? These spaces are regular graphs (i.e.
with each vertex having the same number of incident links)
depending on the particular differential calculus defined on them.

In the present paper we begin an investigation of de Rham
cohomology of finite group manifolds. A systematic analysis is
carried out for finite groups up to order 8.

Table 1 summarizes our findings, and contains the following
informations: name of group, labels of independent one-forms,
number of independent k-forms, Betti numbers.

The alternating sum of Betti numbers always vanishes. Thus the
finite groups up to order 8 have vanishing Euler number (for all
the differential calculi we have considered).

An attempt at self-consistency is made in Section 2, with a
resum\'e on the differential geometry of finite groups. A
graphical representation of the braiding operator and the metric
allows to build a knot invariant for any finite group. Some new
results are also presented in Section 3, where a projector
decomposition of the braiding operator is found, and used to
construct explicitly a projector on the space of 2-forms. The
regular graphs corresponding to particular differential calculi on
$S_3$, $Q$ and $D_4$ are given in Appendix 1.

Section 4 is devoted to de Rham cohomology of finite groups, and
establishes general formulas for the exterior derivative of an
arbitrary $k$-form in terms of matrix $M$ whose kernel yields the
closed $k$-forms. Hodge decomposition theorem holds, the proof
being identical to the one for compact orientable manifolds. Under
some assumptions the Laplacian $\Delta = d\de+\de d$ commutes with
the Hodge operator. Following classical proofs, this implies
Poincar\'e duality.

Section 5 contains some conclusions and open questions.


\sect{Bicovariant calculi on finite groups}


 Let $G$ be a finite group of order $n$ with
generic element $g$ and unit $e$. Consider $Fun(G)$, the set of
complex functions on $G$. An element $f$ of $Fun(G)$ is specified
by its values $f_g \equiv f(g)$ on the group elements $g$, and can
be written as
 \eq
  f=\sum_{g \in G} f_g x^g,~~~f_g \in \mathbb{C} \label{fonG}
 \en
where the functions $x^g$ are defined by
 \eq
  x^g(g') = \de^g_{g'} \label{xg}
\en
Thus $Fun(G)$ is a n-dimensional vector space, and the $n$
functions $x^g$ provide a basis. $Fun(G)$ is also a commutative
algebra, with the usual pointwise sum and product, and unit $I$
defined by $I(g)=1, \forall g \in G$. In particular: \eq x^g
x^{g'}=\de_{g,g'} x^g,~~~\sumong x^g = I \label{mul}
\en
The left and right actions of the group $G$ on itself
 \eq
\lb{dg2a} L_g \, g' = g \, g' = R_{g'} \, g \;\;\; \forall g,g'
\in G \; ,
 \en
induce the left and right actions (pullbacks) ${\cal L}_g$, ${\cal
R}_g$ on $Fun(G)$
 \eq
\lb{dg2b} [{\cal L}_g \, f ] (g') =f( g \, g') = [{\cal R}_{g'} \,
f] (g) \;\;\; \forall f \in Fun(G)\; .
 \en
 For the basis functions we find easily: \eq \Lcal_{g_1} x^{g} =
x^{g_1^{-1} g}, ~~\Rcal_{g_1} x^{g} = x^{g g_1^{-1}}
\en
Moreover:
 \eq \Lcal_{g_1} \Lcal_{g_2}=\Lcal_{g_2g_1},
~~\Rcal_{g_1} \Rcal_{g_2}=\Rcal_{g_1g_2},~~\Lcal_{g_1}
\Rcal_{g_2}=\Rcal_{g_2} \Lcal_{g_1}
 \en
 The $G$ group structure induces a Hopf algebra structure on
$Fun(G)$, and this allows the construction of differential calculi
on $Fun(G)$, according to the techniques of ref.
\cite{wor,ACintro}. We list here the main definitions and
properties. A detailed treatment can be found in \cite{gravfg},
and Hopf algebraic formulas, allowing contact with the general
method of \cite{wor,ACintro}, are listed in the Appendix of
\cite{ACI}.
 \sk
 A (first-order) differential
calculus on $Fun(G)$ is defined by a linear map $d$: $Fun(G)
\rightarrow \Gamma$, satisfying the Leibniz rule $
 d(ab)=(da)b+a(db),~~\forall a,b\in Fun(G)$.
The ``space of 1-forms" $\Ga$ is an appropriate bimodule on
$Fun(G)$, which essentially means that its elements can be
multiplied on the left and on the right by elements of $Fun(G)$.
{} From the Leibniz rule $da=d(Ia)=(dI)a+Ida$ we deduce $dI=0$.
Consider the differentials of the basis functions $x^g$. From
$0=dI=d(\sumong x^g)=\sumong dx^g$ we see that in this calculus
only $n-1$ differentials are independent.
 \sk

A {\sl bicovariant} differential calculus is obtained by requiring
that $\Lcal_g$ and $\Rcal_g$ commute with the exterior derivative
$d$. This requirement in fact {\sl defines} their action on
differentials:
$\Lcal_g db \equiv
 d(\Lcal_g b), \forall b \in Fun(G)$ and similarly for
 $\Rcal_g db$. More generally:
 \eq
 \Lcal_g (adb) \equiv (\Lcal_g a) \Lcal_g db =
 (\Lcal_g a) d (\Lcal_g b)
 \en
 and similarly for $\Rcal_g$.

 \sk
 As in usual Lie group manifolds, we can introduce in $\Ga$
 the left-invariant one-forms $\theta^g$:
 \eq
  \theta^g \equiv
\sumonh x^{hg^{-1}} dx^h =\sumonh x^h dx^{hg}, \label{deftheta}
\en

 It is immediate to check
that indeed $\Lcal_k \theta^g = \theta^g$.
 The right action of $G$ on the elements
$\theta^g$ is given by:
 \eq
  \Rcal_h \theta^g =
\theta^{ad(h)g},~~\forall h \in G \label{Rontheta}
\en
where $ad$ is the adjoint action of $G$ on itself, i.e. $ad(h)g
\equiv hgh^{-1}$. Notice that $\theta^e$ is biinvariant, i.e. both
left and right invariant.

 {} From $\sumong dx^g=0$ one finds: \eq
 \sumong
\theta^g = \sum_{g,h \in G} x^h dx^{hg}= \sumonh x^h \sumong
dx^{hg}=0 \label{sumtheta}
\en
Therefore we can take as basis of the cotangent space $\Ga$ the
$n-1$ linearly independent left-invariant one-forms $\theta^g$
with $g \not= e$. Smaller sets of $\theta^g$ can be consistently
chosen as basis, and correspond to different choices of the
bimodule $\Ga$, see later. Using (\ref{mul}) the relations
(\ref{deftheta}) can be inverted:
 \eq
  dx^h = \sumong x^{hg^{-1}}\theta^g =
   \sumongnote (x^{hg^{-1}} - x^h)\theta^g \label{dxastheta}
\en
 Analogous results hold
for right invariant one-forms $\zeta^g$:
 \eq
  \zeta^g = \sumonh x^{g^{-1}h}dx^h
\en
Using the definition of $\theta^g$ (\ref{deftheta}), the
commutations between $x$ and $\theta$ are easily obtained:
 \eq
  x^h dx^g = x^h \theta^{h^{-1}g} =
\theta^{h^{-1}g} x^g ~~(h\not=g)~~\Rightarrow \theta^g x^h=
x^{hg^{-1}}\theta^g ~~(g\not=e) \label{xthetacomm}
\en
and imply the general commutation rule between functions and
left-invariant one-forms:
 \eq
   \theta^g f = [\Rcal_g f] \theta^g~~~~~~~(g \not= e)
    \label{fthetacomm}
 \en
Thus functions do commute between themselves (i.e. $Fun(G)$ is a
commutative algebra) but do not commute with the basis of
one-forms $\theta^g$. In this sense the differential geometry of
$Fun(G)$ is noncommutative.
 \sk
The differential of an arbitrary function $f \in Fun(G)$ can be
found with the help of (\ref{dxastheta}):
 \eqa
& & d f = \sum_h \, f_h \, d  x^h = \sum_{g,h} f_h \, x^{h \,
g^{-1}} \, \theta^g  = \sum_{g \neq e} \,  ( \sum_h f_h \, x^{h \,
g^{-1}} - f ) \, \theta^g = \nonumber \\ & &~~~~ = \sum_{g \neq e}
\, ( [{\cal R}_g \, f] - f )  \, \theta^g = \sum_{g \neq e} \,  (
t_g \, f) \, \theta^g \; . \label{partflat}
 \ena
\noi Here the finite difference operators  $t_g = \Rcal_g - 1$ are
the analogues of (left-invariant) tangent vectors. They satisfy
the Leibniz rule:
    \eq
 t_g (ff')=
 (t_g f)  f'  +\Rcal_g (f) t_g f'
 =(t_g f) \Rcal_{g} f'  + f t_g f' \label{tgLeibniz}
 \en
 \noi and close on the fusion algebra:
 \eq
 t_g \, t_{g'} = ({\cal R}_{gg'}-1) - ({\cal R}_{g}-1) - ({\cal
R}_{g'}-1) = \sum_h \, C^h_{g,g'} \, t_h \; ,
 \label{fusion}
 \en
  \noi  where the
structure constants $C^h_{g, g'}$ are
 \eq
 C^h_{g,g'} = \delta^h_{g g'} - \delta^h_g - \delta^h_{g'} \; ,
  \label{cconst}
   \en
The commutation rule  (\ref{fthetacomm}) allows to express the
differential of a
 function $f \in Fun(G)$ as a commutator of $f$ with the biinvariant form $\sum_{g \neq
e} \theta^g= -\theta^e $:
 \eq d f = [ \sum_{g \neq e} \theta^g , \, f]= -[\theta^e,f] \; .
 \label{df}
 \en
 An {\sl exterior product}, compatible with the left and right
actions of $G$, can be defined as
 \eqa & &
 \theta^g \wedge \theta^{g'}
 = \theta^g \otimes \theta^{g'} - \sum_{k,k'} \Lambda^{g \, g'}_{~~k'
\, k}  \theta^{k'} \otimes \theta^k = \theta^g \otimes \theta^{g'}
-  \theta^{g g' g^{-1}} \otimes \theta^g = \nonumber\\
 & & ~~~~~~ = \theta^g
\otimes \theta^{g'} - [{\cal R}_g \theta^{g'} ] \otimes \theta^g
\; , \;\;\; (g,g' \neq e) \; , \label{extheta}
 \ena
 where the tensor product between elements $\rho,\rhop \in \Ga$ is
defined to have the properties $\rho a\otimes \rhop=\rho \otimes a
\rhop$, $a(\rho \otimes \rhop)=(a\rho) \otimes \rhop$ and $(\rho
\otimes \rhop)a=\rho \otimes (\rhop a)$. The braiding matrix
$\La$:
 \eq
\La^{g \, g'}_{~~k' \, k} = \delta^{g g' g^{-1}}_{k'} \,
\delta^g_k \; ,~~~~\Linv{g \, g'}{k' \, k} =
\delta^{g'}_{k'}~\delta^{g'^{-1} g g'}_{k} \,~~ \;\;\; (g,g' \neq
e) \; .
 \en
 satisfies the Yang-Baxter equation $\Lambda^{nm}_{~~ij} \Lambda^{js}_{~~kq}
 \Lambda^{ik}_{~~rp}
 = \Lambda^{ms}_{~~kj} \Lambda^{nk}_{~~ri} \Lambda^{ij}_{~~pq}$ (or in condensed notation $\La_{12} \La_{23} \La_{12}=
  \La_{23} \La_{12} \La_{23}$). With this exterior product we find
 \eq
\theta^g \wedge \theta^g = 0 \;\;\; (\forall g) \; , \;\;\;
\theta^g \wedge \theta^{g'} = - \theta^{g'} \wedge \theta^{g}
\;\;\; (\forall g,g': \;\; [g, \, g']=0 \; , \;\; g \neq e) \; .
 \en
 Left and right actions on $\Ga \otimes \Ga$ are
  simply defined by:
  \eq
  \Lcal_h (\rho \otimes \rhop)= \Lcal_h \rho \otimes \Lcal_h
  \rhop,~~~
\Rcal_h (\rho \otimes \rhop)= \Rcal_h \rho \otimes \Rcal_h
  \rhop
  \en
 Compatibility  of the exterior product with $\Lcal$ and $\Rcal$
 means that
 \eq
 \Lcal(\theta^i \we \theta^j)=\Lcal\theta^i \we \Lcal
 \theta^j, ~~\Rcal(\theta^i \we \theta^j)=\Rcal\theta^i \we \Rcal
 \theta^j
 \en
 Only the second relation is nontrivial and is verified upon use
 of the definition (\ref{extheta}).
 We can generalize the previous definition to
 exterior products of $k$ left-invariant one-forms:
 \eq
\theta^{i_1} \we ... \we \theta^{i_k} \equiv
A^{i_1..i_k}_{j_1..j_k}~
 \theta^{j_1} \otimes ...\otimes
\theta^{j_k} \label{multiwedge}
\en
\noi or in short-hand notation:
  \eq \theta^{1} \we ... \we
\theta^{k}= A_{1...k}~
 \theta^{1} \otimes ...\otimes
\theta^{k}
\en
\noi The labels $1...k$ in $A$ refer to index couples, and
$A_{1,...k}$ is the analogue of the antisymmetrizer of $k$ spaces,
defined by the recursion relation
 \eq
  A_{1\ldots k} = [1 - \Lambda_{k-1,k} +
\Lambda_{k-2,k-1} \Lambda_{k-1,k}-   \ldots -(-1)^k \Lambda_{12}
\Lambda_{23} \cdots \Lambda_{k-1,k}] A_{1\ldots k-1} ,
\label{Asymm}
\en
 where $A_{12} = 1-\La_{12}$.
  The space of $k$-forms $\Ga^{\we k}$ is therefore
defined as in the usual case but with the new permutation operator
$\La$, and can be shown to be a bicovariant bimodule (see for ex.
\cite{Athesis}), with left and right action defined as for $\Ga
\otimes ...\otimes \Ga$ with the tensor product replaced by the
wedge product. The graded bimodule $\Omega = \sum_k \Ga^{\we k}$,
with $\Ga^{\we k}=\Ga^{\otimes k}/Ker(A_{1...k})$, is the exterior
algebra of forms.

 The {\sl exterior derivative} is
defined as a linear map $ d~:~\Gamma^{\we k} \rightarrow
\Gamma^{\we (k+1)}$ satisfying $d^2=0$ and the graded Leibniz rule
 \eq
 d(\rho \we \rhop)=d\rho \we \rhop +
(-1)^k \rho \we d\rhop \label{propd1}
\en
\noi where $\rho \in \Ga^{\we k}$, $\rhop \in \Ga^{\we k'}$,
$\Ga^{\we 0} \equiv Fun(G)$ . Left and right action is defined as
usual:
 \eq
 \Lcal_g (d\rho)=d \Lcal_g \rho, ~~~\Rcal_g (d\rho)=d \Rcal_g \rho
 \label{propd2}
\en
In view of relation (\ref{Rontheta}) the algebra $\Omega$ has
natural quotients over the ideals $H_g$= $\{ \theta^{hgh^{-1}},
\forall h \}$, corresponding to the various conjugacy classes of
the elements $g$ in $G$. The different bicovariant calculi on
$Fun(G)$ are in 1-1 correspondence with different quotients of
$\Omega$ by any sum of the ideals $H=\sum H_g$, cf.
\cite{DMGcalculus,FI1,gravfg}. In practice one simply sets
$\theta^g = 0$ for all $g \not= e$ not belonging to the particular
union $G'$ of conjugacy classes characterizing the differential
calculus. The dimension of the space of independent 1-forms for
each bicovariant calculus on $Fun(G)$ is therefore equal to the
dimension of the subspace $\Ga/H$. If there are $r$ nontrivial
conjugacy classes, the number of possible unions $G'$ of these
classes is $2^r-1$. We have then $2^r-1$ differential calculi.
 \sk
 The {\sl Cartan-Maurer equation} for the differential forms $\theta^g$
(\ref{deftheta}) is obtained by direct calculation, using the
definition (\ref{deftheta}), the expression  (\ref{dxastheta}) of
$dx^h$ in terms of $\theta$'s, and the commutations
(\ref{xthetacomm}):
 \eq
   d \theta^g = - \sum_{h \not= e,h'\not= e} \de^g_{hh'} \theta^h \we \theta^{h'} +
 \sum_{k \not= e} \theta^k \we \theta^g + \sum_{k \not= e} \theta^g \we \theta^k =
- \sum_{h \neq e} \sum_{h' \neq e} \, C^g_{h,h'} \,  \theta^h
\wedge \theta^{h'} \; ,~~(g\not=e) \label{CM}
 \en
  where the structure constants $C^g_{h,h'}$ are given in (\ref{cconst}).
Using  the identity:
 \eq
  \sum_{h \not= e,h'\not= e} \, \delta^k_{hh'}
\, \theta^h \wedge \theta^{h'} = \sum_{h \not= e,h'\not= e} \,
\delta^k_{hh'} \, \left( \theta^h \otimes \theta^{h'}-
 \theta^{h h'h^{-1}} \otimes \theta^h \right) = 0 \label{iden}
 \en
 \noi the Cartan-Maurer equation can be rewritten by means of the
anticommutator of $\theta^g$ with the biinvariant form $\theta^e$:
 \eq
 d\theta^g=-\theta^e \we \theta^g - \theta^g \we \theta^e
 \label{dtheta}
 \en
 cf. the case of 0-forms (\ref{df}). Considering now a generic
 element $\rho = a \theta$ of $\Gamma$ it is easy to find that
 $d\rho = -\theta^e \we \rho - \rho \we \theta^e$. The general
 rule is
 \eq
 d\rho = [-\theta^e, \rho]_{grad} \equiv
 -\theta^e \we \rho + (-1)^{deg(\rho)} \rho \we \theta^e
 \en
 valid for any $k$-form, where $[-\theta^e, \rho]_{grad}$ is the
 graded commutator.
 \sk
 There are two (Hopf
algebra) {\sl conjugations} on $Fun(G)$ \cite{DMGcalculus,CasPag}
 \eq
(x^g)^* = x^g \; , \;\;\; (x^g)^\star = x^{g^{-1}}
 \label{conj}
 \en
These involutions can be extended to the whole exterior (Hopf)
algebra $\Omega$:
 \eq
  (\theta^g)^*= - \theta^{g^{-1}} \; ,
  \;\;\; (\theta^g)^\star = \zeta^{g} \;
   \en
such that $(\rho \we \rhop)^* = (-1)^{deg(\rho) deg(\rhop)}
\rhop^* \we \rho^*$ etc. We'll use the *-conjugation in the
sequel. Consistency of this conjugation
 requires that if $\theta^g \neq 0$ then $\theta^{g^{-1}} \neq 0$ as well:
 we have to include in $\Ga/H$ at least the two
ideals $H_g$ and $H_{g^{-1}}$ (if they do not coincide). We obtain
thus a ${}^*$-differential calculus, i.e. $(df)^*=d ( f^*)$.
 \sk
 In fact the conjugations can also be defined directly on the
 tensor algebra. For example $(\theta_1 \otimes \theta_2)^*=\La(\theta^*_2 \otimes
 \theta^*_1)$, or with explicit indices
 $(\theta^{i_1} \otimes \theta^{i_2})^*= \theta^{i^{-1}_2 i^{-1}_1 i_2} \otimes
  \theta^{i^{-1}_2}$. This rule is consistent with $(\theta_1 \we
\theta_2)^*=
 - \theta^*_2 \we \theta^*_1$ as one proves by recalling that
 $[ \La (\theta_1 \otimes \theta_2)]^* = \La^{-1}[ (\theta_1 \otimes
\theta_2)^*]$. In general:
 \eqa
 & & (\theta^{i_1} \otimes \theta^{i_2} \otimes \cdots \otimes
 \theta^{i_k})^*= (-1)^k~
 \theta^{ad(i_2...i_k)^{-1} i^{-1}_1} \otimes
 \theta^{ad(i_3...i_k)^{-1} i_2^{-1}} \otimes \cdots\otimes
 \theta^{i_k^{-1}} \label{startensor} \\
 & & (\theta^{i_1} \we \theta^{i_2} \we \cdots \we
 \theta^{i_k})^*= (-1)^{k(k+1)\over 2}~\theta^{i_k^{-1}}\we \cdots \we \theta^{i_2^{-1}}
 \we \theta^{i_1^{-1}} \label{starwedge}
  \ena

 \sk
  The fact that both $\theta^g$ and $\theta^{g^{-1}}$ are included in
  the basis of left-invariant 1-forms characterizing
   the differential calculus also ensures the existence of a unique metric
    (up to a normalization).

  The {\sl metric} is defined as a bimodule pairing, symmetric
  on left-invariant 1-forms. It maps couples of 1-forms
  $\rho,\sigma$ into $Fun(G)$, and satisfies the properties
 \eq
  <f\rho,\sigma h>=f<\rho,\sigma>h~,~~<\rho f,\sigma>
  =<\rho, f\sigma>~. \label{metricproperties}
 \en
 where $f$ and $h$ are arbitrary functions belonging to $Fun(G)$.
 Up to a normalization the above properties
 determine the metric on the left-invariant 1-forms. Indeed from
 $<\theta^g,f\theta^h>=<\theta^g,\theta^h>\Rcal_{h^{-1}}f= \Rcal_g
 f <\theta^g,\theta^h>$ one deduces:
 \eq
 g^{rs} \equiv <\theta^r,\theta^s> \equiv -\de^r_{s^{-1}} \label{metric}
 \en
 the minus sign being a convenient choice of normalization
 (so that (\ref{dualitypairing}), and consequently the
 positivity property of (\ref{pospairing}) holds).
 Thus $g^{rs}$ is symmetric and $\theta^r$ has nonzero
  pairing only with $\theta^{r^{-1}}$.
 The pairing is compatible with the ${}^*$-conjugation
  \eq
 <\rho,\sigma >^*=<\sigma^*,\rho^*> \label{starmetric}
  \en

 We can generalize $<\!\!~,\!\!~>$  to tensor products
 of left-invariant $1$-forms as follows (as proposed in the second ref. of \cite{ACI}):
 \eq
 < \theta^{i_1} \otimes \cdots \otimes \theta^{i_k},
  \theta^{j_1} \otimes \cdots \otimes \theta^{j_k}> \equiv
  g^{i_k,j_k} g^{i_{k-1},ad(i_k)j_{k-1}} g^{i_{k-2},ad(i_{k-1} i_k) j_{k-2}}
   ...~g^{i_1,ad(i_2 \cdots i_k)j_1} \label{pairingtensors}
  \en
  Using (\ref{startensor}) we find the duality relation:
  \eq
   < (\theta^{i_1} \otimes \cdots \otimes \theta^{i_k})^*,
  \theta^{i_1} \otimes \cdots \otimes \theta^{i_k}>=1
  \label{dualitypairing}
  \en

  The pairing (\ref{pairingtensors}) is extended to all tensor
  products by $ <f\rho,\sigma h>=f<\rho,\sigma>h$ where now $\rho$ and $\sigma$ are
  generic tensor products of same order. Then we prove easily
  that $<\rho f,\sigma> =<\rho, f\sigma>$ for any function $f$, so
  that $<\!\!~,\!\!~>$ is a bimodule pairing. Moreover
  $<\rho,\sigma >=<\sigma,\rho>$, i.e. the pairing is symmetric, but only
   when $\rho$ and $\sigma$ are tensor products of
  $\theta$'s,  as one can prove from the
  definition (\ref{pairingtensors}). Another interesting property
  is
  \eq
  < \rho, \rho^* > = N(\rho)~ |f|^2 \label{pospairing}
  \en
  where $\rho$ is a generic $k$-form
   $\rho=f~ \theta^{i_1} \we ... \we \theta^{i_k}$ and $N(\rho)$ is a
   real positive constant
   depending on $\rho$.
   For example $<\theta^{i_1}
  \we \theta^{i_2},(\theta^{i_1}
  \we \theta^{i_2})^*> = 2$ (in this case $N(\rho)$ does not
  depend on $\rho$).

 \sk

In general for a differential calculus with $m$ independent
tangent vectors, there is an integer $p  \geq m$ such that the
linear space of left-invariant $p$-forms is 1-dimensional, and
$(p+1)$- forms vanish identically \footnote{with the exception of
$Z_2$, see ref. \cite{tmr}}. This is so far an experimental
result, based on the examples we have studied. It implies that
every product of $p$ basis one-forms $\theta^{g_1} \we
\theta^{g_2} \we ... \we \theta^{g_p}$ is proportional to one of
these products, which can be chosen to define the volume form
$vol$:
 \eq
 \theta^{g_1} \we \theta^{g_2} \we ... \we \theta^{g_p}=
 \epsilon^{g_1,g_2,...g_p}~ vol \label{vol}
 \en
 where $\epsilon^{g_1,g_2,...g_p} $ is the proportionality constant.
 The volume $p$-form is obviously left invariant. It is also right invariant
 \cite{gravfg} (the proof is based on the $ad(G)$ invariance of
 the $\epsilon$ tensor:
 $\epsilon^{ad(g)h_1,...ad(g)h_p}=\epsilon^{h_1,...h_p}$).
 \sk
 Finally, if $vol = \theta^{k_1} \we ...\we \theta^{k_p}$, then
 \eq
 vol^*=(-1)^{p (p+1) \over 2}\epsilon^{k_p^{-1}...k_1^{-1}} vol
 \en
 so that $vol$ is either real or imaginary. If $vol^* = -vol$  we can always
 multiply it by $i$ and obtain a real volume form. In that case
 comparing $(\theta^{g_1} \we ...\we \theta^{g_p})^*
 =(-1)^{p(p+1)\over 2} \theta^{g_p^{-1}}\we ... \we \theta^{g_1^{-1}} =
 \epsilon^{g_p^{-1}...g_1^{-1}} (-1)^{p(p+1)\over 2} vol$
 with $(\theta^{g_1} \we ...\we \theta^{g_p})^* =
 \epsilon^{g_1...g_p} ~(vol)^* =  \epsilon^{g_1...g_p} ~vol$
 yields
 \eq
 \epsilon^{g_p^{-1}...g_1^{-1}} = (-1)^{p(p+1)\over 2} ~\epsilon^{g_1...g_p}
 \label{epsiminus}
 \en
 The pairing of the volume with itself is simply:
 \eq
 <vol,vol> = N(vol)
 \en

 \sk
Having identified the volume $p$-form it is natural to define the
integral of a function on $G$
 \eq
 \int f ~vol  = \sumong f(g) \label{intpform}
 \en
 \noi the right-hand side being just the Haar measure of the function
$f$.

 Due to the biinvariance of the volume form, the integral map $\int
: \Ga^{\we p} \mapsto \mathbb{C}$ satisfies the biinvariance
conditions:
 \eq
  \int \Lcal_g \rho = \int \rho = \int \Rcal_g \rho
  \en

  Moreover, under the assumption that
  $d(\theta^{g_2} \we ... \we \theta^{g_p})=0$, i.e.
  that any exterior product of $p-1$ left-invariant one-forms $\theta$ is closed,
 the important property holds:
  \eq
  \int df =0
  \en
  with $f$  any $(p-1)$-form: $f=f_{g_2,...g_p}~ \theta^{g_2}
\we ... \we \theta^{g_p}$. This property, which allows
  integration by parts, has a simple proof (see ref. \cite{gravfg}).
  When the volume form belongs to a nontrivial
  cohomology class, $d(\theta^{g_2} \we ... \we \theta^{g_p})$ must vanish
  (otherwise it should be proportional to $vol$, and this
  contradicts $vol \not= d\rho$) and therefore integration by
  parts holds.
  \sk

 The {\sl Hodge dual}, an important ingredient for gauge theories,
 has been defined in \cite{AB,ACI} as the unique map from $k$-forms $\sigma$ to $(p-k)$-forms
 $* \sigma$ such that
\begin{equation}
\label{defhodge} \rho\wedge *\sigma = <\rho,\sigma> vol
~~~~~~~~~\rho,\sigma  {\mbox{ $k$-forms}}
\end{equation}
The Hodge dual is left linear; if $vol$ is central it is also
right linear :
 \eq *(f\rho\,h)=f(*\rho)h
\label{linearityhodge}
 \en
 with $f,h \in Fun(G)$.
 Moreover
 \eq
 {}* 1= 1~vol~~,~~*vol=N(vol)~~ \label{hodge1vol}
 \en
  \noi {\bf Conjecture 1:} the  definition (\ref{defhodge}) is equivalent to
  the following explicit expression of the Hodge dual on the
  exterior products of $\theta$ 1-forms:
  \eq
   *(\theta^{i_1} \we ... \we \theta^{i_k})= const \cdot
    \epsilon_{j_{k+1}...j_p}^{~~~~~~~~~i_1
  ...i_k} ~~\theta^{j_p} \we ...\we \theta^{j_{k+1}}
  \label{defhodge2}
  \en
  for an appropriate value of $const$, and where
  the $j$ indices of the epsilon tensor are lowered
  by means of the metric $g_{i,j}$. We can easily
  verify a necessary condition for the equivalence: setting $\rho =
  \theta^{i_1} \we ... \we \theta^{i_k}$ into
  \eq
  \rho^* \we *\rho=<\rho^*,\rho>~vol=N(\rho) ~vol \label{rhorho}
  \en
  is indeed consistent with (\ref{defhodge2}) because of
  (\ref{epsiminus}).
  \sk
  \noi {\bf Conjecture 2:} Hodge duality is an involution.
  It is so when acting on $0$-forms and on $vol$: indeed (\ref{hodge1vol})
  imply $**1=N(vol)$, $**vol=N(vol)~vol$. When acting on a generic
  $k$-form, the Hodge duality being an involution is consistent
  with (\ref{rhorho}) (although it is not clear that it is implied by
  it). Indeed, the conjugate of (\ref{rhorho}) is:
   \eq
   (*\rho)^* \we \rho = <\rho^*,\rho> vol
   \en
   On the other hand, substituting $\rho \rightarrow *\rho$
   into (\ref{rhorho}) yields
   \eq
   (*\rho)^* \we **\rho = <(*\rho)^*,*\rho> vol
   \en
   These two relations are consistent with
   \eq
   **\rho = { <(*\rho)^*,*\rho> \over <\rho^*,\rho>} \rho
   \en
   i.e. with the involutive property of $*$.
   \sk

  In the case of the 3-D calculus on $S_3$ a
  Hodge involution $**=id$ can be defined on the basis $k$-forms as in
  (\ref{defhodge2}) (see also the second ref. in \cite{majidetal}) with
 $const=1/\sqrt{3}$.
 \sk
 \noi {\bf Note 1:}
 the ``group manifold" of a finite group is simply a
  collection of points corresponding to the
group elements, linked together in various ways, each
corresponding to a particular differential calculus on $Fun(G)$
\cite{DMGcalculus,gravfg}. The links are associated to the tangent
vectors $\Rcal_h -1$ of the differential calculus, or equivalently
to the right actions $\Rcal_h$, where $h$ belongs to the union
$G'$ of conjugacy classes characterizing the differential
calculus. Two points $x^g$ and $x^{g'}$ are linked if
$x^{g'}=\Rcal_h x^g$, i.e. if $g'=gh^{-1}$  for some $h$ in $G'$.
The link is oriented from $x^g$ to $x^{g'}$ (unless $h=h^{-1}$ in
which case the link is unoriented): the resulting ``manifold" is
an oriented graph. From every point exactly $m$ (= number of
independent 1-forms) links originate. Appendix 1 contains the
graphs for differential calculi on finite groups up to order
eight.

\sk \noi {\bf Note 2: Knot invariants}.
 \sk
 We can represent the
braiding operator $\La$ and its inverse $\La^{-1}$ as

\begin{center}
\begin{picture}(300,50)
   \put(0,0){\begin{picture}(100,50)
              \put(0,0){$\La^{ab}_{~~cd}~=$}
              \thicklines
              \put(60,-20){\begin{picture}(50,50)
                             \put(10,10){\line(1,1){12}}
                             \put(10,40){\line(1,-1){30}}
                             \put(40,40){\line(-1,-1){12}}
                             \put(0,0){c} \put(0,42){a}
                             \put(40,0){ d} \put(40,42){ b}
                           \end{picture}}
             \end{picture}}

   \put(150,0){\begin{picture}(100,50)
              \put(0,0){$\La^{-1~ab}_{~~~~~~cd}~=$}
              \thicklines
              \put(60,-20){\begin{picture}(50,50)
                             \put(10,10){\line(1,1){30}}
                             \put(10,40){\line(1,-1){12}}
                             \put(40,10){\line(-1,1){12}}
                             \put(0,0){c} \put(0,42){a}
                             \put(40,0){ d} \put(40,42){ b}
                           \end{picture}}
             \end{picture}}
\end{picture}
\end{center}

\sk\sk

\noi The metric $g_{ab}$ is represented as

\begin{center}
\begin{picture}(300,50)
   \put(0,0){\begin{picture}(100,50)
              \put(0,0){$g_{ab}~=$}
              \thicklines
              \put(60,-10){\begin{picture}(50,50)
                             \put(25,20){\oval(25,25)[t]}
                              \put(10,10){$a$}
                              \put(35,10){$b$}
                          \end{picture}}
             \end{picture}}

   \put(150,0){\begin{picture}(100,50)
              \put(0,0){$g^{ab}~=$}
              \thicklines
              \put(60,-10){\begin{picture}(50,50)
                             \put(25,20){\oval(25,25)[b]}
                              \put(10,25){$a$}
                              \put(35,25){$b$}
                           \end{picture}}
             \end{picture}}
\end{picture}
\end{center}

 \sk
  \noi  The metric $g_{ab}$ allows to close the braids into
knots, and the above graphical representations yield a knot
invariant for any finite group. This invariant is an integer
number KN.
 The three Reidemeister moves hold because of

i) $g_{ab} \La^{ab}_{~~cd} = g_{cd}$,

ii) the definition of the crossings corresponding to $\La$ and
$\La^{-1}$,

iii) the Yang-Baxter equations for $\La$ and the properties:

\eq
 g_{ab} ~\La^{bc}_{~~de}=\La^{-1~cb}_{~~~~~~ad} ~g_{be}
\en

\noi Thus the unknot has KN equal to $g^{ab} g_{ab}=m$, the
dimension of the differential calculus. The KN of the right
trefoil is:
 \eq
 \Linv{a_1a_2}{~~a_3a_4} \Linv{b_1b_2}{~~b_3b_4} \Lambda^{a_4b_3}_{~~~~c_3c_4}
 ~g^{c_3a_3} g^{c_4b_4} ~g_{a_2b_1} g_{a_1b_2}
 \en

\noi the left trefoil KN being obtained by $\La \leftrightarrow
\La^{-1}$. Up to finite groups of order eight, the KN does not
distinguish between left and right trefoils, and its values are
given in Table 1.

\let\picnaturalsize=N
\def\picsize{0.7in}
\def\picfilename{trefoil.eps}
\ifx\nopictures Y\else{\ifx\epsfloaded Y\else\input epsf \fi
\let\epsfloaded=Y
\centerline{\ifx\picnaturalsize N\epsfxsize \picsize\fi
\epsfbox{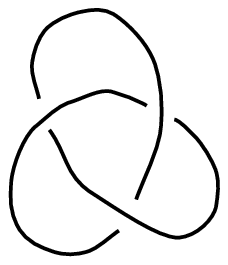}}}\fi
\sk
 \centerline{ {\small{\bf Fig. 1} : right trefoil}}
 \sk

\sk \noi {\bf Note 3: Summary of conventions}.
 \sk
 $n$ : order of the finite
group $G$.

$m$ : number of independent 1-forms, depends on the particular
differential calculus.

$k$ : generic rank of a form.

$p$ : rank of top forms (volume).

  \sect{$k$-forms, components and projector decompositions}

The nontriviality of the braiding operator $\La$ entails some
complication in the analysis of the space of $k$-forms. Its
dimension is usually larger than
$(\stackrel{k}{{}_{\scriptstyle{m}}})$ as when 1-forms simply
anticommute. For example the space of two-forms in the case S3 (3D
calculus) is four-dimensional, and not three-dimensional as it
would be in ordinary differential geometry for a 3D-manifold. The
basis $\Theta^I_{(k)}$ of $k$-forms is determined by finding the
null eigenvectors of the generalized antisymmetrizer $A^i_j$
defined in (\ref{multiwedge}), $i,j$ being here composite indices
$i=(i_1...i_k)$ etc. Suppose there are $q$ independent null
eigenvectors. Then the space of $k$ forms must have dimension
$m^k-q$, since $A^i_j$ maps the $m^k$ - dimensional space of $k$ -
tensors to the space of $k$-forms. The null eigenvectors lead to a
set of independent linear relations between $k$-forms: we can
solve these relations in terms of the basis elements
$\Theta^I_{(k)}$, {\it \small{I}}~{$=1,...(m^k-q)$. When their
number is not too large, these basis elements are given in
Appendix 1.

Any $k$-form $B$ can be expanded on the $\Theta^I_{(k)}$ basis:
$B=B_I~\Theta^I_{(k)}$, where $B_I$ are the components of $B$ on
the basis. It may be of some interest to retain explicit
information on the $\theta^{i_1} \we...\we \theta^{i_k}$ structure
of the basis when extracting components. This could be useful, for
example, when defining the analogue of Riemann curvature and its
contractions, as in ref.s \cite{gravfg}-\cite{DMgrouplattice}.

Consider the 2-form $B_{ij} ~\theta^i \we \theta^j$. What we need
is really a {\sl projector} such that
 \eq
 {\cal A}^{ij}_{~~kl} ~\theta^k \we \theta^l = \theta^i \we \theta^j
 \label{projectorA}
 \en
 Then the components of $B$ can be extracted as
 \eq
  {\cal A}^{ij}_{~~kl}~B_{ij}
  \en

 The generalized antisymmetrizer $A = id - \La$ is a projector only
 when $\La = \La^{-1}$, i.e. when it is really an antisymmetrizer.
 To find the projector ${\cal A}$ the key observation is that there always
 exists a power $s$ such that \cite{DMGcalculus}
 \eq
 \La^s = id
 \en
 In fact this $s$ is given by  $s=2 |ad(G)|$, where $|ad(G)|$ denotes the number of
 elements of  $ad(G):=\{ad(g)|g\in G\}$, the group of inner
 automorphisms of $G$.

 We recall the proof of  \cite{DMGcalculus} : for any $a \in ad(G)$ let
 $C(a)$ denote the cyclic subgroup of $ad(G)$ generated by $a$.
 Since $ad(G)$ is a finite group, $|C(a)|$ is finite and
 $a^{|C(a)|}=id$. Furthermore, $|C(a)|$ is a divisor of $|ad(G)|$
 by Lagrange theorem. Finally, notice that from
 \eq
  \La (\theta^g \otimes \theta^h) = \theta^{ad(g)h} \otimes
  \theta^g
  \en
  one finds by induction
  \eqa
  & & \La^{2k-1}  (\theta^g \otimes \theta^h) = \theta^{ad(gh)^k h}
  \otimes \theta^{ad(gh)^{k-1} g} \\
  & & \La^{2k}  (\theta^g \otimes \theta^h) = \theta^{ad(gh)^k g}
  \otimes \theta^{ad(gh)^{k} h}
  \ena
 From the last equation the proof follows.
 \sk
 Defining the {\sl order} of $\La$ to be the smallest positive
 integer $s$ such that $\La^s=id$, the previous proof implies that
 $s \leq 2 |ad(G)|$. In general the equality does not hold. For
 example, the symmetric groups $S_n$ with $n > 3$ (and universal calculus)
 have $s < 2|ad(G)|$ \cite{DMGcalculus}.
 \sk
 Next we notice that $\La^s = id$ means that the eigenvalues of
 $\La$ are the $s-th$ roots of unity, i.e. $(1,q,q^2,...q^{s-1})$
 with $q=e^{2\pi i\over s}$. Then, if we denote by $P_i$ the
 projector on the eigenspace corresponding to the root $q^i$, the
 braiding operator has the projector decomposition:
 \eq
  \La = P_0 + q P_1 + q^2 P_{2} +...q^{s-1} P_{s-1}
  \en
 Using the projector properties $\sum_0^{s-1} P_i = id$ and  $P_i P_j = \delta_{ij} P_i$ yields
 the system of $s$ operator equations:
 \eqa
 & & id = P_0 + P_1 + P_2 +...+ P_{s-1} \nonumber \\
 & & \La = P_0 + q P_1 + q^2 P_2 +...+ q^{s-1} P_{s-1}\nonumber \\
 & & \La^2 = P_0 + q^2 P_1 + q^4 P_2 +...+ q^{2(s-1)} P_{s-1}
 \nonumber\\
 & & \vdots \nonumber\\
 & & \La^{s-1}=\La^{-1} = P_0 + q^{-1} P_1 + q^{-2} P_2 +...+ q P_{s-1}
 \ena
 Using $1+q+q^2+...+q^{s-1}=0$ these can be easily inverted:
 \eq
 P_i = {1\over s} \sum_{j=0}^{s-1} q^{-ij}\La^j
 \en
 and one can check directly the projector properties.
 \sk
 \noi The projector $P_0$ satisfies the relation:
 \eq
 P_0 ~(id-\La)=0~~\Rightarrow P_0~ \theta^i \we \theta^j= 0
 \label{P0}
 \en
 since $P_0 \La = P_0$. On the other hand the complementary projector
 \eq
 id - P_0 =  id-{1\over s} [id + \La + \La^2
 +... \La^{s-1}]={1\over s}[(id-\La)+(id-\La^2) + ...(id-\La^{s-1})]
 \en
 applied to $\theta^i \we \theta^j$ leaves it unvaried. Then
 ${\cal A}=id-P_0$ is the projector on two-forms we were looking for, satisfying
 (\ref{projectorA}).
 \sk
 Notice that the components $B_{ij}$ as defined by $B=B_{ij}
 \theta^i \we \theta^j$ are ambiguous, since
 \eq
 B_{ij} \rightarrow  B_{ij}+c_{kl}~(P_0)^{kl}_{~~ij},~~~c_{kl} \in
 Fun(G)
 \en
 correspond to the same 2-form $B$ (use (\ref{P0}). This ambiguity
 is fixed by projecting with ${\cal A}$: the projection removes
 any piece in $B_{ij}$ proportional to $(P_0)^{kl}_{~~ij}$ due to
 $ {\cal A} P_0 = (id-P_0)P_0 = 0$.


\sect{De Rham cohomology}


\subsection{Cohomology classes}

Cohomology classes are found by computing the null vectors of the
linear mapping $d^{(k)}: \Ga^{\we k} \rightarrow \Ga^{\we (k+1)}$
(exterior derivative acting on $k$-forms). These give the closed
forms in $\Ga^{\we k}$, and the exact forms in $\Ga^{\we (k+1)}$
(as the image of the space orthogonal to the closed forms in
$\Ga^{\we k}$). As usual, the number of independent closed but not
exact $k$-forms is simply the difference between
$dim[Ker(d^{(k)})]$ and $dim[Im(d^{(k-1)})]$.

These numbers, i.e. the Betti numbers, as well as the explicit
list of cohomology representatives, can be computed by
 finding the null vectors of the
  matrix $M$ representing $d^{(k)}$.
 \sk
 Let us determine this matrix in terms of quantities related to
 the differential calculus on $Fun(G)$. A generic
 $k$-form $B$ can be expanded on the basis of $k$-forms
 $\Theta_{(k)}^I$: $B =B_I(x) \Theta_{(k)}^I$. Moreover, its
 components being functions on $G$, can be themselves expanded on
 the basis $x^{g}$ defined in (\ref{xg}): $B_I (x) = B_{Ig}x^g$.
 By means of the definitions:
 \eqa
  & & d \Theta^I_{(k)} = C^I_{J} ~ \Theta^J_{(k+1)} \label{defC}\\
  & & \theta^i \we  \Theta^I_{(k)} = T^{iI}_J ~ \Theta^J_{(k+1)}
  \label{defT}
  \ena
  the exterior derivative on the generic $k$-form $B$ becomes:
  \eqa
  & & dB = (dB_I) \we \Theta^I_{(k)} + B_I C^I_J~ \Theta^J_{(k+1)}=
    [(R_i-1)B_I]~\theta^i \we \Theta^I_{(k)}+B_I C^I_J ~\Theta^J_{(k+1)}
    \nonumber\\
    & & ~~~~=  [(R_i-1)B_I~T^{iI}_{J}+B_I C^I_J]
    ~\Theta^J_{(k+1)}= [(R_i-1)(B_{Ig}x^g)~T^{iI}_{J}+B_{Ig} x^g~ C^I_J]
    ~\Theta^J_{(k+1)} \nonumber\\
    & & ~~~~= [(B_{Ig}x^{gi^{-1}}-B_{Ig} x^g)~T^{iI}_{J}+B_{Ig}x^g ~C^I_J]
    ~\Theta^J_{(k+1)}
    \ena
 Projecting on the bases $x^{g'}$ and $\Theta^J_{(k+1)}$ yields
 finally:
 \eq
  [dB]_{Jg'} = M_{Jg'}^{~~~Ig}~ B_{Ig}
  \en
  with
  \eq
   M_{Jg'}^{~~~Ig} = \sum_{i }T^{iI}_{J}~(\de^g_{g'i} -
   \de^g_{g'})+C^I_J~\de^g_{g'}
   \en
   This matrix has  $dim(G) \times dim(\Ga^{\we (k+1)})$ rows and
    $dim(G) \times dim(\Ga^{\we (k)})$ columns. The quantities $T^{iI}_{J}$
    and $C^I_J$ defined in (\ref{defC}), (\ref{defT}) are easily obtained
     from the Cartan-Maurer equations (and the Leibniz rule),
      and the expansion of $k+1$ forms $\theta^{i_1} \we ...\we \theta^{i_{k+1}}$
       on the basis $\Theta^J_{(k+1)}$.

     Suppose that $M_{Jg'}^{~~~Ig}$  has
    $q$ null eigenvectors $V^{\al}$, $\al=1,...q$ with components
    $V^{\al}_{Ig}$. Then there are $q$ independent closed
    $k$-forms $C_{(k)}^{\al}$
    given by:
    \eq
    C_{(k)}^{\al} = V^{\al}_{Ig}~x^{g} ~\Theta^I_{(k)},~~~\al=1,...q
    \en
   This analysis has been carried out for all finite groups up to
   order 8, and the results are summarized in the following Table.

   \sk\sk
  {\bf Table 1: de Rham cohomology of $S_3$, $Q$, $D_4$, $Z_N$ }
  \sk

 \begin{tabular}{|c|}
   \hline\hline
   $S_3$~~~~~~$\theta^a,\theta^b,\theta^c$ ~~~~~~~~~~~~~~~~~~~
   ~~~~~~~~~~~~~~~~~~~~~~~~~~~~~~~~~~~~~~~~~~~~~~~~~~~~~~~~~~~~~~KN=9
    \\ \hline
 \end{tabular}

 \begin{tabular}{|c|c|c|c|c|c|}
   order  & 0 & 1 & 2 & 3 & 4 \\
   $\sharp$  & 1 & 3 & 4 & 3 & 1 \\
    $b_k$ & 1 & 1 & 0 & 1 & 1 \\ \hline
 \end{tabular}

  \begin{tabular}{|c|}
   \hline
   $S_3$~~~~~~$\theta^{ab},\theta^{ba}$ ~~~~~~~~~~~~~~~~~~~~~~~~~~~~
    ~~~~~~~~~~~~~~~~~~~~~~~~~~~~~~~~~~~~~~~~~~~~~~~~~~~~~~~KN=2
    \\ \hline
 \end{tabular}

 \begin{tabular}{|c|c|c|c|c|c|}
   order & 0 & 1 & 2  \\
   $\sharp$ & 1 & 2 & 1  \\
   $b_k$ & 2 & 4 & 2 \\ \hline
 \end{tabular}

  \begin{tabular}{|c|}
   \hline
   $S_3$~~~~~~$\theta^a,\theta^b,\theta^c,\theta^{ab},\theta^{ba}$ ~~~~~~
   ~~~~~~~~~~~~~~~~~~~~~~~~~~~~~~~~~~~~~~~~~~~~~~~~~~~~~~~~~~~~~~~~KN=11
    \\ \hline
 \end{tabular}

 \begin{tabular}{|c|c|c|c|c|c|c|c|c|}
   order  & 0 & 1 & 2 & 3 & 4 & 5 & 6 & 7 \\
   $\sharp$  & 1 & 5 & 14 & 31 & 58 & 95 & 140 & ... \\
    $b_k$ & 1 & 2 & 1 & 2 & 4 & ... & ... & ... \\ \hline
 \end{tabular}

 \begin{tabular}{|c|}
   \hline
   $Q$~~~~~~$\theta^i,\theta^{i^{-1}},\theta^j,\theta^{j^{-1}}$ ~~~~~~~~~~~~~~~
   ~~~~~~~~~~~~~~~~~~~~~~~~~~~~~~~~~~~~~~~~~~~~~~~~~~~~~~~~~~~KN=4
    \\ \hline
 \end{tabular}

 \begin{tabular}{|c|c|c|c|c|c|c|c|c|c|}
   order  & 0 & 1 & 2 & 3 & 4 & 5 & 6 & 7 & 8 \\
   $\sharp$  & 1 & 4 & 8 & 12 & 14 & 12 & 8 & 4 & 1 \\
    $b_k$ & 1 & 2 & 1 & 2 & 4 & 2 & 1 & 2 & 1 \\ \hline
 \end{tabular}

 \begin{tabular}{|c|}
   \hline
   $D_4$~~~~~~$\theta^2,\theta^4,\theta^5,\theta^6$ ~~~~~~~~~~~~~~~~~~
   ~~~~~~~~~~~~~~~~~~~~~~~~~~~~~~~~~~~~~~~~~~~~~~~~~~~~~~~~~~~KN=4
    \\ \hline
 \end{tabular}

 \begin{tabular}{|c|c|c|c|c|c|c|c|c|c|}
   order  & 0 & 1 & 2 & 3 & 4 & 5 & 6 & 7 & 8 \\
   $\sharp$  & 1 & 4 & 8 & 12 & 14 & 12 & 8 & 4 & 1 \\
    $b_k$ & 1 & 2 & 1 & 2 & 4 & 2 & 1 & 2 & 1 \\ \hline
 \end{tabular}

 \begin{tabular}{|c|}
   \hline
   $Z_N$~~~~~~$\theta^{u},\theta^{u^{-1}}$ ~~~~~~~~~~~~~~~~~~~~~~~~~~~~
    ~~~~~~~~~~~~~~~~~~~~~~~~~~~~~~~~~~~~~~~~~~~~~~~~~~~~~KN=2
    \\ \hline
 \end{tabular}

 \begin{tabular}{|c|c|c|c|c|c|}
   order & 0 & 1 & 2  \\
   $\sharp$ & 1 & 2 & 1  \\
   $b_k$ & 1 & 2 & 1 \\ \hline
 \end{tabular}
 \sk\sk
 \noi where the order $k$ of independent forms, the number $\sharp$
 of independent $k$-forms and the $k$-th Betti
 number $b_k$ are given for the three nonabelian groups
 $S_3,Q,D_4$ and for the cyclic groups $Z_N$. We give only partial results for
 the universal calculus on $S_3$, the volume form being of order at least 12.
 The independent
 one-forms characterizing the differential calculus are also
 indicated (see the Appendix for conventions), together with the
 knot numbers KN for the trefoils.

 \subsection{Adjoint, Laplacian and Poincar\'e duality}

 We first define an inner product between two generic $k$-forms
  as follows:
 \eq
 \langle\langle \rho,\sigma \rangle\rangle \equiv \int_G <\rho^*,\sigma> vol~= \int_G \rho^*
 \we (*\sigma)
 \en
 This product is positive definite because of (\ref{pospairing}).
 It can be extended to the direct sum ${\bigoplus}_k\Gamma^{\wedge
k}$, requiring the spaces $\Gamma^{\wedge k}$ and $\Gamma^{\wedge
k'}$ to be orthogonal if $k\neq k'$.
 As usual, we define the adjoint of the exterior
 derivative as the unique mapping
 $\delta:\Gamma^{\wedge k}\longrightarrow \Gamma^{\wedge(k-1)}$ such that

 \eq
  \langle\langle d\alpha,\beta\rangle\rangle =\langle\langle
 \alpha,\delta \beta\rangle\rangle,~~\forall
 \alpha\in\Gamma^{\wedge\left(  k-1\right) },~~ \forall
 \beta\in\Gamma^{\wedge k}. \label{adjointd}
 \en

{\sl Lemma 1:} if $\int d \rho =0$, $\forall$ (p-1)-form $\rho$
(see Sect. 2) then:
 \eq
 d * = (-1)^k * \delta
 \en

 {\sl Proof:} let $\al$, $\be$ be generic $k-1$ and $k$-forms
 respectively. Then
 \eq
 d(\al^* \we *\be)=d\al^* \we *\be + (-1)^{k-1} \al^* \we d(*\be)
 \en
 Integrating on the group, using $\int d =0$ and (\ref{adjointd}) yields
 \eq
 \int \al^* \we *\delta \be = (-1)^k \int \al^* \we d(*\be)
 \en
 which implies the theorem, since
 $\langle\langle~,~\rangle\rangle$ is positive definite.
 \sk
  Suppose now that $** = \eta ~id$ (Conjecture 2 of
 Sect.2), where $\eta$ is a sign. Then
 \eq
 \de = (-1)^k \eta *d*~~~~\Longrightarrow ~~~~*\Delta = \Delta *
 \en
 where
 \eq
 \Delta \equiv d\de + \de d
 \en
 is the Laplacian. The commutation of the Laplacian with the Hodge operator
 allows to reproduce the standard proof for Poincar\'e duality, so
 that
 \eq
 dim(H^k) = dim(H^{p-k})
 \en

 Note that the Hodge decomposition theorem holds in any case, the proof relying on
 the finiteness of the space of harmonic $k$-forms. Then every
 cohomology class contains a unique harmonic representative.

\sect{Conclusions and outlook}

We have started an investigation on the (de Rham) cohomological
properties of finite groups. Most of the classical results for
differential manifolds can be translated into this setting, since
they are based on algebraic relations holding also for finite
groups. A challenging question for future work is how to relate de
Rham cohomology of finite groups to the homology of the regular
graphs that encode their differential calculi.

Although we have not discussed it in the present paper, a parallel
transport commuting with the left and right action of the finite
group can be introduced, as well as a torsion and a curvature.
This allows the construction of Yang-Mills, Born-Infeld and
gravity actions on finite groups, as mentioned in the
Introduction. It would be of interest to find how cohomology
information (for example the analogue of characteristic classes)
reflects itself on the dynamics of these theories.

\app{Differential calculi on finite groups of order $\leq 8$}

\subsection{ The permutation group $S_3$}

 \sk
 Elements: $a=(12)$, $b=(23)$, $c=(13)$, $ab=(132)$, $ba=(123)$, $e$.
 \sk
 Multiplication table:
 \sk
 \begin{tabular}{|c|c|c|c|c|c|c|}
   \hline
     & e & a & b & c & ab & ba \\  \hline
   e & e & a & b & c & ab & ba \\   \hline
   a & a & e & ab & ba & b & c \\ \hline
   b & b & ba & e & ab & c & a \\ \hline
   c & c & ab & ba & e & a & b \\ \hline
   ab & ab & c & a & b & ba & e \\ \hline
   ba & ba & b & c & a & e & ab \\ \hline
 \end{tabular}
 \sk
 Nontrivial conjugation classes: $I = [a,b,c]$, $II = [ab,ba]$.
 \sk
There are 3 bicovariant calculi $BC_I$, $BC_{II}$, $BC_{I+II}$
corresponding to the possible unions of the conjugation classes.
They have respectively dimension 3, 2 and 5.

 \subsubsection{ $BC_I$ differential calculus}

Basis of the 3-dimensional vector space of one-forms:
 \eq
\theta^a,~\theta^b,~\theta^c
\en
We'll use the shorthand notation $\{i_1,...i_k \} = \theta^{i_1}
\we ...\theta^{i_k}$.

 \noi Basis $\Theta_{(2)}$ of the 4-dimensional vector space of
two-forms:
 \eq
  \{a,b\},~
 \{b,c\},~\{a,c\},~\{c,b\}
 \en

 Any other wedge product of two $\theta$'s can be expressed
 as linear combination of the basis elements:
 \eq
  \{b,a\} = -\{a,c\} - \{c,b\},~~\{c,a\}=-\{a,b\}-\{b,c\}
  \en

\noi Basis $\Theta_{(3)}$ of the 3-dimensional vector space of
three-forms:
 \eq
 \{a,b,c\},~\{a,c,b\},
  ~\{b,a,c\}
  \en

 and:
  \eqa
 & &\{c,b,a\}=- \{c,a,c\}=-\{a,c,a\}=\{a,b,c\} \nonumber \\
 & &\{b,c,a\}=- \{b,a,b\}=-\{a,b,a\}=\{a,c,b\} \nonumber \\
 & &\{c,a,b\}=- \{c,b,c\}=-\{b,c,b\}=\{b,a,c\} \ena

\noi Basis $\Theta_{(4)}$ of the 1-dimensional vector space of
four-forms:

\eq
 vol = \{a,b,a,c\}
 \en

 \noi The $\epsilon$ tensor is defined by:
  \eq
 \{g_1,g_2,g_3,g_4\}=
 \epsilon^{g_1,g_2,g_3,g_4}~ vol \label{epsiI}
 \en
 \noi Its nonvanishing components are:
 \eqa
 & &
 \epsilon^{abac}=\epsilon^{acab}=\epsilon^{cbca}=\epsilon^{cacb}=
 \epsilon^{babc}=\epsilon^{bcba}=1 \\
 & &
 \epsilon^{baca}=\epsilon^{caba}=\epsilon^{abcb}=\epsilon^{cbab}=
 \epsilon^{acbc}=\epsilon^{bcac}=-1 \label{epsivaluesI}
 \ena
 Note the centrality of $vol$:
 \eq
 f ~vol = vol~f,~~~\forall f \in Fun(G) \label{fvol}
 \en
 due to $\Rcal_a \Rcal_b \Rcal_a \Rcal_c = \Rcal_{abac}=\Rcal_e = id$
\sk
 \noi Cartan-Maurer equations:
\eqa
 & & d\theta^a=-\theta^b\we\theta^c-\theta^c \we \theta^b
 \nonumber \\
 & & d\theta^b=-\theta^a\we\theta^c+\theta^a \we \theta^b +
 \theta^b \we \theta^c
 \nonumber \\
 & & d\theta^c = -\theta^a\we\theta^b+ \theta^a \we \theta^c +
 \theta^c \we \theta^b
 \ena

 The exterior derivative on any three-form
of the type $\theta \we \theta \we \theta$ vanishes, as one can
easily check by using the Cartan-Maurer equations and the
equalities between exterior products given above. Equivalently,
the volume form belongs to a nontrivial cohomology class ($H^4$).
Then, as discussed in Section 2, integration of a total
differential vanishes on the ``group manifold" of $S_3$
corresponding to the $BC_I$ bicovariant calculus. This ``group
manifold" has three independent directions, associated to the
cotangent basis $\theta^a,~\theta^b,~\theta^c$. Note however that
the volume element is of order four in the left-invariant
one-forms $\theta$. \sk

\noi De Rham cohomology (generators):
 \sk
 \eq
 H^0=I,~~H^1= X,~~H^2=0,~~H^3=  (*X)
 ,~~H^4 = ~vol
 \en
 where $X=\theta^a+\theta^b+\theta^c$

 \subsection{ $BC_{II}$ differential calculus}

Basis of the 2-dimensional vector space of one-forms:
 \eq
\theta^{ab},~\theta^{ba}
\en
\noi Basis of the 1-dimensional vector space of two-forms:
 \eq
  vol = \{ab,ba\}=-\{ba,ab\}
 \en
so that:
  \eq
 \{g_1,g_2\}=
 \epsilon^{g_1,g_2} vol \label{epsiII}
 \en
 where the $\epsilon$ tensor is the usual 2-dimensional Levi-Civita tensor.
 Again $f~vol = vol~f$ since $abba=e$.
 \sk
\noi Cartan-Maurer equations: \eq
 d\theta^{ab} =0,~~ d\theta^{ba} =0
 \en

 Thus the exterior derivative on any one-form $\theta^g$
 vanishes and integration of a total differential
vanishes on the group manifold of $S_3$ corresponding to the
$BC_{II}$ bicovariant calculus. This group manifold has two
independent directions, associated to the cotangent basis
$\theta^{ab},~\theta^{ba}$.
 \sk

\noi De Rham cohomology:
 \eqa
 & & H^0=(x^a+x^b+x^c)~I,~~~~ (x^e+x^{ab}+x^{ba})~I,\\
 & & H^1= (x^a+x^b+x^c)~ \theta^{ab},~~
   (x^e+x^{ab}+x^{ba})~ \theta^{ab},\\
  & & ~~~~~~~~~  (x^a+x^b+x^c)~ \theta^{ba} ,~~
   (x^e+x^{ab}+x^{ba}) ~\theta^{ba},\\
  & & H^2=
  (x^a+x^b+x^c)~vol,~~(x^e+x^{ab}+x^{ba})~vol.
 \ena

\subsubsection{The $S_3$ group ``manifold"}

 We can draw a picture of the group manifold of $S_3$. It is made
 out of 6 points, corresponding to the group elements and identified with
 the functions $x^e,x^a,x^b,x^c,x^{ab},x^{ba}$.
\sk $BC_I$ - calculus: \sk
 From each
 of the six points $x^g$ one can move in three directions, associated to
 the tangent vectors $t_a,t_b,t_c$, reaching three other points
 whose ``coordinates" are
 \eq
\Rcal_a x^g = x^{ga},~~\Rcal_b x^g = x^{gb},~~\Rcal_c x^g = x^{gc}
\en
 The 6 points and the ``moves" along the 3 directions are
 illustrated in the Fig. 2. The links are not oriented since
 the three group elements $a,b,c$ coincide with their inverses.

 \sk
$BC_{II}$ - calculus: \sk
 From each
 of the six points $x^g$ one can move in two directions, associated to
 the tangent vectors $t_{ab},t_{ba}$, reaching two other points
 whose ``coordinates" are
 \eq
\Rcal_{ab} x^g = x^{gba},~~\Rcal_{ba} x^g = x^{gab}
\en
 The 6 points and the ``moves" along the 3 directions are
 illustrated in Fig. 2. The arrow convention on a link
 labeled  (in italic) by a group element $h$
 is as follows: one
 moves in the direction of the arrow via the action
 of $\Rcal_{h}$ on $x^g$. (In this case $h=ab$). To move in the opposite
 direction just take the inverse of $h$.
\sk
 \noi Note that the $BC_{II}$ graph has two disconnected pieces.
 This explains $b_0 \equiv dim(H^0)=2$.

\let\picnaturalsize=N
\def\picsize{4.0in}
\def\picfilename{S3.eps}
\ifx\nopictures Y\else{\ifx\epsfloaded Y\else\input epsf \fi
\let\epsfloaded=Y
\centerline{\ifx\picnaturalsize N\epsfxsize \picsize\fi
\epsfbox{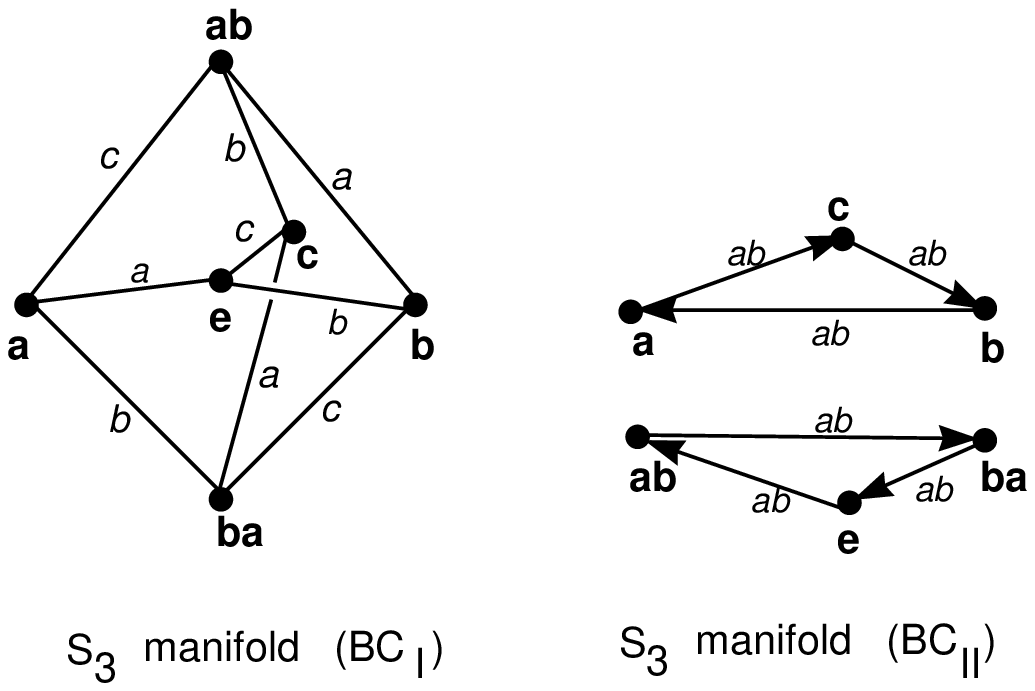}}}\fi

 \sk
  {\small{\bf Fig. 2} : $S_3$ group manifold, and moves of the
   points under the group action}
 \sk

\subsection{The quaternion group $Q$}

\noi Elements of $Q$: $\{e,~-e, ~i,~-i, ~j, ~-j,~k, ~-k\} $
 \sk
 \noi Multiplication table: $ij=k$ and cyclic, $i^2 = -e$ etc.
 \sk
 \noi Nontrivial conjugation classes:

 $[-e]=\{ -e \} ~;~ [i]=\{i,-i \}
~;~ [j]=\{j,-j \} ~;~ [k]=\{k,-k \} ~;~$
 \sk
 There are differential calculi of dimensions 1 up to 7
 (universal calculus). Many are isomorphic. The 1D, 2D, 3D differential
  calculi are rather trivial. We give here details
 on a 4-dimensional calculus corresponding to the union of the $[i]$
 and $[j]$ conjugation classes.

 \subsubsection{4D-differential calculus}

 \noi Basis of 1-forms:
  $\theta^i,\theta^{i^{-1}},\theta^j,\theta^{j^{-1}}.$
   \sk
   \noi Basis of 2-forms:

   $\{-i,i\},\{-i,j\},\{-i,-j\},\{j,i\},\{j,-i\},\{-j,i\},\{-j,-i\},\{-j,j\}.$
   \sk
   and

   $\{i,-i\}=-\{-i,i\},~~\{j,-j\}=-\{-j,j\},$

   $\{i,j\}=-\{j,-i\}-\{-i,-j\}-\{-j,i\},~~\{i,-j\}=-\{j,i\}-\{-i,j\}-\{-j,-i\}.$
 \sk
 \noi Basis of 3-forms:

 $\{-i,j,i\},\{-i,j,-i\},\{-i,-j,i\},\{-i,-j,-i\},\{-i,-j,j\},
 \{j,i,-i\},$

 $\{j,-i,-j\},\{-j,-i,i\},\{-j,-i,j\},\{-j,-i,-j\},
 \{-j,j,i\},\{-j,j,-i\}.$
  \sk

  \noi Basis of 4-forms:

  $\{-i,j,-i,i\},\{-i,-j,-i,i\},\{-i,-j,-i,j\},\{-i,-j,j,i\},\{-i,-j,j,-i\},$

 $\{j,i,-i,j\}, \{j,-i,-j,-i\},\{-j,-i,j,i\},\{-j,-i,j,-i\},\{-j,-i,-j,i\},$

 $\{-j,-i,-j,-i\}, \{-j,-i,-j,j\},\{-j,j,-i,i\}, \{-j,j,-i,-j\}.$

  \sk
 \noi Basis of 5-forms:

 $\{-i,j,-i,-j,i\},\{-i,-j,-i,j,i\},\{-i,-j,-i,j,-i\},\{-i,-j,j,-i,i\},$

 $\{-i,-j,j,-i,-j\}, \{-j,-i,j,-i,i\}, \{j,-i,j,-i,-j\},\{-j,-i,-j,-i,i\},$

 $\{-j,-i,-j,-i,j\},\{-j,-i,-j,j,i\},\{-j,-i,-j,j,-i\}, \{-j,j,-i,-j,i\}.$

 \sk
 \noi Basis of 6-forms:

  $\{-i,-j,-i,j,-i,i\},\{-i,-j,-i,j,-i,-j\},\{-i,-j,j,-i,-j,i\},$

  $\{-j,-i,j,-i,-j,i\}, \{-j,-i,-j,-i,j,i\}, \{-j,-i,-j,-i,j,-i\},$

  $\{-j,-i,-j,j,-i,i\},\{-j,-i,-j,j,-i,-j\}.$

   \sk
 \noi Basis of 7-forms:

 $ \{-i, -j, -i, j, -i, -j, i \}, \{-j, -i, -j, -i, j, -i, i\},$

 $ \{-j, -i, -j, j, -i, -j, i \},    \{ -j, j, -i, -j, -i, j, -i \} $

 \sk

 \noi The volume form $vol = \{-j,-i,-j,-i,j,-i,-j,i \}$ is central.

 \sk
 \noi The epsilon tensor has 928 nonvanishing components, with values
 $1,-1,2,-2$ (mostly $1,-1$).
 \sk

 \noi Cartan-Maurer equations:

  $ d \theta^i= -\theta^j \we \theta^{-i}-\theta^{-i} \we \theta^j-
  \theta^{-i} \we \theta^{-j} - \theta^{-j} \we \theta^{-i}$

  $ d \theta^{i^{-1}}=\theta^j \we \theta^{-i}+\theta^{-i} \we
  \theta^j+
  \theta^{-i} \we \theta^{-j} + \theta^{-j} \we \theta^{-i}$

  $ d \theta^j = \theta^j \we \theta^{i}+\theta^{-i} \we \theta^j-
  \theta^{-i} \we \theta^{-j} - \theta^{-j} \we \theta^{i}$

  $ d \theta^{j^{-1}} =-\theta^j \we \theta^{i}-\theta^{-i} \we
  \theta^j+
  \theta^{-i} \we \theta^{-j} + \theta^{-j} \we \theta^{i} $

  \sk

  \noi De Rham cohomology:

  $ H^0:~I, $

  $ H^1:~ X^i= \theta^i + \theta^{-i},~X^j=\theta^j +
  \theta^{-j},$

  $H^2:~
  X^i\we X^j,$

  $H^3:~ W=\theta^{-i} \we \theta^{j} \we \theta^{-i} +
  \theta^{-i} \we \theta^{-j} \we \theta^{-i},~~Z=
  \theta^{-i} \we \theta^{j} \we \theta^i +
  \theta^{-j} \we \theta^{-i} \we \theta^{i},$

   $H^4:~X^i \we W,~~ X^j \we W, ~~X^j \we Z, ~~
   \theta^{-j}\we\theta^j\we\theta^{-i}\we\theta^{-j} +
   \theta^{-j}\we\theta^{-i}\we\theta^{-j}\we\theta^{j}$

  \sk with $X^i \we X^j = - X^j \we X^i, X^i \we
  X^i = X^j \we X^j =0$.

\let\picnaturalsize=N
\def\picsize{2.5in}
\def\picfilename{quat4.eps}
\ifx\nopictures Y\else{\ifx\epsfloaded Y\else\input epsf \fi
\let\epsfloaded=Y
\centerline{\ifx\picnaturalsize N\epsfxsize \picsize\fi
\epsfbox{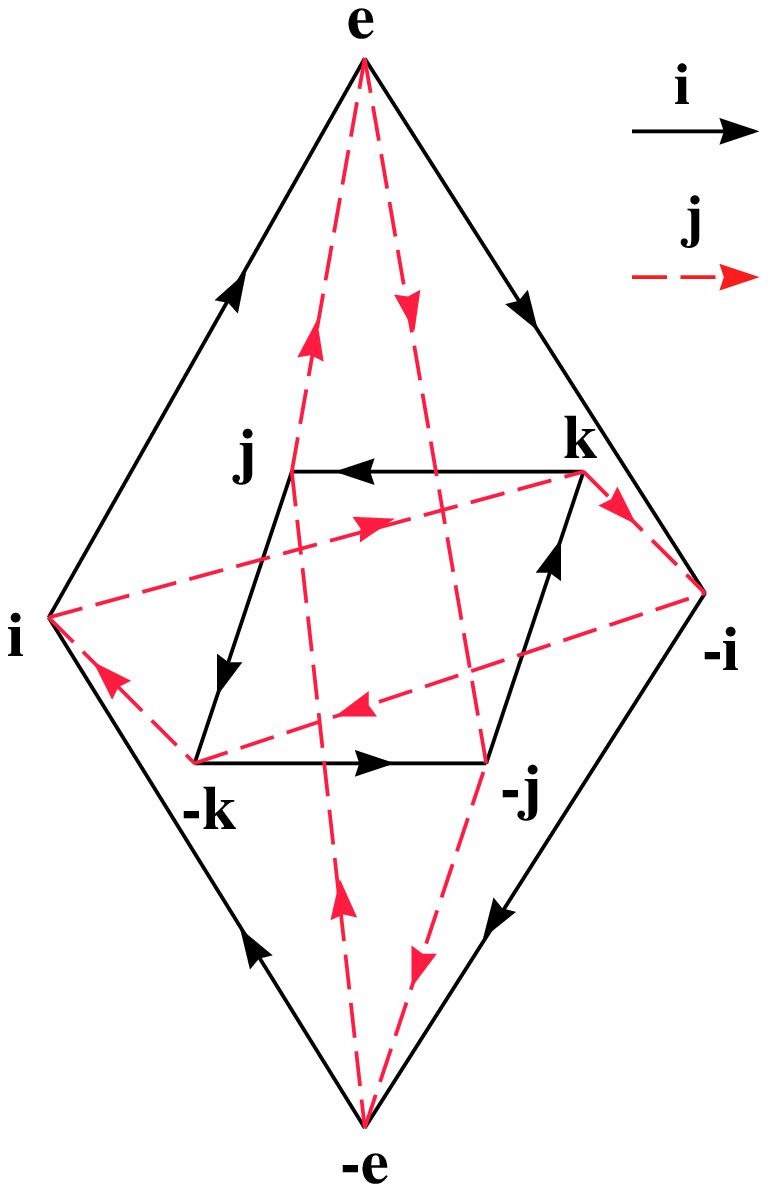}}}\fi

 \sk
  {\small{\bf Fig. 3} : $Q$ group manifold corresponding to the [i,-i,j,-j]
  differential calculus}
 \sk

  \subsection{Dihedral group $D_4$}

  $D_4$ : group of isometries of the square ABCD.
  \sk

  \noi Elements of $D_4$:
  \sk

  $1$ = identity $e$

  $2$ = ${\pi\over 2}$ clockwise rotation

  $3$ = (diag AC) (diag BD)

  $4$ = ${\pi\over 2}$ anticlockwise rotation

  $5$ = horizontal reflection

  $6$ = vertical reflection

  $7$ = (diag BD)

  $8$ = (diag AC)
  \sk
  where (diag AC) and (diag BD) are the reflections on the two
  diagonals.
 \sk
  \noi Multiplication table:
  \sk
\begin{tabular}{|c|c|c|c|c|c|c|c|c|}
  \hline
  . & 1 & 2 & 3 & 4 & 5 & 6 & 7 & 8 \\
  \hline
  1 & 1 & 2 & 3 & 4 & 5 & 6 & 7 & 8 \\
  \hline
  2 & 2 & 3 & 4 & 1 & 8 & 7 & 5 & 6 \\
  \hline
  3 & 3 & 4 & 1 & 2 & 6 & 5 & 8 & 7 \\
  \hline
  4 & 4 & 1 & 2 & 3 & 7 & 8 & 6 & 5 \\
  \hline
  5 & 5 & 7 & 6 & 8 & 1 & 3 & 2 & 4 \\
  \hline
  6 & 6 & 8 & 5 & 7 & 3 & 1 & 4 & 2 \\
  \hline
  7 & 7 & 6 & 8 & 5 & 4 & 2 & 1 & 3 \\
  \hline
  8 & 8 & 5 & 7 & 6 & 2 & 4 & 3 & 1 \\ \hline
\end{tabular}
 \sk

 \sk
 \noi Nontrivial conjugation classes: $[3], [2,4], [5,6], [7,8]$.
 There are 15 bicovariant calculi.

 \subsubsection{4D-differential calculus}

 \noi Basis of 1-forms:
  $\theta^2,\theta^{4},\theta^5,\theta^{6}.$
   \sk
   \noi Basis of 2-forms:

  $\{4,2\},\{4,5\},\{4,6\},\{5,2\},\{5,4\},\{6,2\},\{6,4\},\{6,5\}$
   \sk
   and

  $\{2,4\}=-\{4,2\},\{5,6\}=-\{6,5\},$

  $\{2,5\}=-\{4,6\} - \{5,4\}-\{6,2\},\{2,6\}=-\{4,5\} -
  \{5,2\}-\{6,4\}$

 \sk
 \noi Basis of 3-forms:

 $\{4,5,2\},\{4,5,4\},\{4,6,2\},\{4,6,4\},\{5,4,2\},\{5,4,5\}$

 $\{5,4,6\},
 \{6,4,2\},\{6,4,5\},\{6,4,6\},\{6,5,2\},\{6,5,4\}$

  \sk

  \noi Basis of 4-forms:

   $\{4,5,4,2\},\{4,6,4,2\},\{4,6,4,5\},\{5,4,5,2\},\{5,4,5,4\},\{5,4,6,2\},\{5,4,6,4\}$

 $\{6,4,5,2\},
 \{6,4,5,4\},\{6,4,6,2\},\{6,4,6,4\},\{6,5,4,2\},\{6,5,4,5\},\{6,5,4,6\}$

 \sk

  \noi Basis of 5-forms:

   $  \{4, 6, 4, 5, 2 \}, \{4, 6, 4, 5, 4\} ,  \{5, 4, 5, 4, 2\}, \{5, 4, 6, 4, 2\},
      \{5, 4, 6, 4, 5\},  \{6, 4, 5, 4, 2\} $

      $ \{6, 4, 6, 4, 2\} , \{6, 4, 6, 4, 5\}, \{6, 5, 4, 5, 2\} , \{6, 5, 4, 5, 4\} ,
        \{6, 5, 4, 6, 2\}, \{6, 5, 4, 6, 4\} $

 \sk

 \noi Basis of 6-forms:

   $\{4, 6, 4, 5, 4, 2\} , \{5, 4, 6, 4, 5, 2\}, \{5, 4, 6, 4, 5, 4\},  \{6, 4, 6, 4, 5,2\},$

   $ \{6, 4, 6, 4, 5, 4\} , \{6, 5, 4, 5, 4, 2\} ,\{6, 5, 4, 6, 4, 2\} , \{6, 5, 4, 6, 4,5 \} $
 \sk
  \noi Basis of 7-forms:

  $ \{5, 4, 6, 4, 5, 4, 2\},  \{6, 4, 6, 4, 5, 4, 2\}, \{6, 5, 4, 6, 4, 5, 2\},
  \{6, 5, 4, 6, 4, 5, 4 \}$

  \sk

   \noi The volume form $\{6,5,4,6,4,5,4,2 \}$ is central.

   \sk

 \noi The epsilon tensor has 928 nonvanishing components, with values
 $1,-1,2,-2$ (mostly $1,-1$). Note the perfect similarity with the
 quaternion case.
 \sk\sk

 \noi Cartan-Maurer equations:

  $ d \theta^2=-\theta^4 \we \theta^5 -\theta^4 \we \theta^6 -
    \theta^5 \we \theta^4 -\theta^6 \we \theta^4 $

  $ d \theta^{4}=\theta^4 \we \theta^5 +\theta^4 \we \theta^6 +
    \theta^5 \we \theta^4 +\theta^6 \we \theta^4$

  $ d \theta^5 = \theta^4 \we \theta^5 -\theta^4 \we \theta^6 +
    \theta^5 \we \theta^2 -\theta^6 \we \theta^2$

  $ d \theta^{6} = -\theta^4 \we \theta^5 +\theta^4 \we \theta^6 -
    \theta^5 \we \theta^2 +\theta^6 \we \theta^2$

  \sk

  \noi De Rham cohomology:

  $H^0:~ I,$

  $H^1:~X=\theta^2 + \theta^4,~Y=\theta^5 + \theta^6,$

  $H^2=X \we Y,$

  $H^3:~W=\{4,5,4\}+\{4,6,4\},~~Z=\{4,5,2\}+\{6,4,2\}+\{4,6,2\}+\{5,4,2\},$

  $H^4= X\we W,~~Y\we W,~~Y \we Z,~~Z \we X$

  \sk

  with  $X \we
  Y = - Y \we X, ~X \we X = Y \we Y =0$.

\let\picnaturalsize=N
\def\picsize{2.5in}
\def\picfilename{d4.eps}
\ifx\nopictures Y\else{\ifx\epsfloaded Y\else\input epsf \fi
\let\epsfloaded=Y
\centerline{\ifx\picnaturalsize N\epsfxsize \picsize\fi
\epsfbox{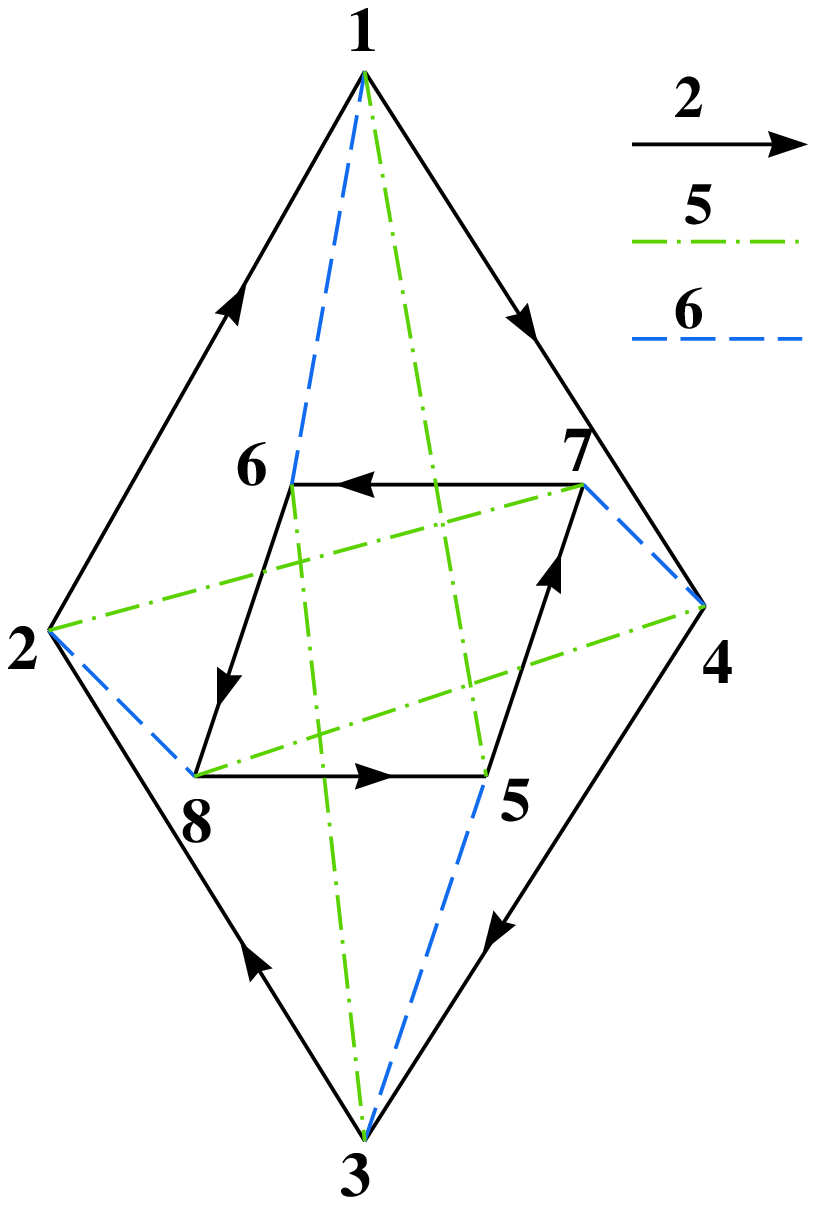}}}\fi

 \sk
  {\small{\bf Fig. 4} : $D_4$ group manifold,corresponding to the
  [2,4,5,6] differential calculus}
 \sk


\end{document}